\documentclass[conference, 9pt]{IEEEtran}
\IEEEoverridecommandlockouts
\usepackage{amsmath,amssymb,amsfonts}
\usepackage{circuitikz}
\usepackage{subcaption}
\usepackage{tikz}
\usetikzlibrary{decorations.pathreplacing,calligraphy}
\usepackage{booktabs}
\usepackage{dcolumn}
\newcolumntype{d}[1]{D{.}{.}{#1}}
\usepackage{pgf}
\usepackage{pgfplots}
\usetikzlibrary{positioning}
\usepackage{pifont}
\usepackage{adjustbox}
\usepackage[nolist, nohyperlinks]{acronym}
\usepackage{nicefrac}
\usepackage{hyperref}
\usepackage{cleveref}
\usepackage{stmaryrd}
\usepackage{setspace}
\usepackage{flushend}

\acrodefplural{nn}[NNs]{neural networks}
\acrodefplural{smm}[SMMs]{spatial mixture models}
\acrodefplural{irm}[IRMs]{ideal ratio masks}
\acrodefplural{ibm}[IBMs]{ideal binary masks}
\acrodefplural{doa}[DoAs]{directions of arrival}

\begin{acronym}
\acro{stft}[STFT]{short-time Fourier transform}
\acro{rir}[RIR]{room impulse response}
\acro{istft}[iSTFT]{inverse short-time Fourier transform}
\acro{doa}[DoA]{direction of arrival}
\acro{irm}[IRM]{ideal ratio mask}
\acro{mvdr}[MVDR]{minimum-variance distortionless response}
\acro{nn}[NN]{neural network}
\acro{dnn}[DNN]{deep neural network}
\acro{gcc_phat}[GCC-PHAT]{generalized cross-correlation phase transform}
\acro{vad}[VAD]{voice activity detection}
\acro{tdoa}[TdoA]{time difference of arrival}
\acro{ssl}[SSL]{sound source localization}
\acro{srp_phat}[SRP-PHAT]{steered response power with phase transform}
\acro{slc}[SLC]{spatial likelihood coding}
\acro{sslc}[SSLC]{spatio-spectral likelihood coding}
\acro{mse}[MSE]{mean squared error}
\acro{mae}[MAE]{mean absolute error}
\acro{tcn}[TCN]{time-convolutional network}
\acro{sisdr}[SI-SDR]{scale-invariant signal-to-distortion ratio}
\acro{pesq}[PESQ]{perceptual evaluation of speech quality}
\acro{cacgmm}[cACGMM]{complex angular central gaussian mixture model}
\acro{wer}[WER]{word error rate}
\acro{pit}[PIT]{permutation-invariant training}
\acro{lbt}[LBT]{location-based training}
\acro{ssm}[SSM]{spatial mixture model}
\acro{asr}[ASR]{automatic speech recognition}
\acro{ds}[DS]{direct separation}
\acro{dc}[DC]{deep clustering}
\acro{scc}[SCC]{spatial classification coding}
\acro{sscc}[SSCC]{spatio-spectral classification coding}
\acro{lboc}[LBOC]{location-based output channels}
\acro{lbol}[LBOL]{location-based output likelihood}
\acro{bce}[BCE]{binary cross-entropy}
\acro{src}[SRC]{spatial regression coding}
\acro{ssrc}[SSRC]{spatio-spectral regression coding}
\acro{jnf}[JNF]{joint non-linear filter}
\acro{hac}[HAC]{hierachical agglomerative clustering}
\acro{sbc}[SBC]{spatial binary coding}
\acro{slc}[SLC]{spatial likelihood coding}
\acro{mwsbc}[MW-SBC]{mask-weighted spatial binary coding}
\acro{mwslc}[MW-SLC]{mask-weighted spatial likelihood coding}
\acro{nn}[NN]{neural network}
\acro{dnn}[DNN]{deep neural network}
\acro{fbmest}[FB-MEst]{full-band mask estimator}
\acro{mccruse}[MC-CRUSE]{multi-channel convolutional recurrent U-net architecture for speech enhancement}
\acro{estoi}[ESTOI]{extended short-time objective intelligibility}
\acro{snr}[SNR]{signal-to-noise ratio}
\end{acronym}

\def\x{{\mathbf x}}

\def\channel{C}

\definecolor{tab_blue}{RGB}{31,119,180}
\crefname{figure}{Fig.}{Figs.}  %
\Crefname{figure}{Figure}{Figures}    %
\usepackage{cite}
\usepackage{algorithmic}
\usepackage{graphicx}
\usepackage{textcomp}
\usepackage{xcolor}
\usepackage{orcidlink}
\def\BibTeX{{\rm B\kern-.05em{\sc i\kern-.025em b}\kern-.08em
    T\kern-.1667em\lower.7ex\hbox{E}\kern-.125emX}}
\begin{document}

\title{Mask-Weighted Spatial Likelihood Coding for Speaker-Independent Joint Localization and Mask Estimation}

\author{\IEEEauthorblockN{Jakob Kienegger \orcidlink{0009-0003-0926-8142}}
\IEEEauthorblockA{\textit{Signal Processing Group} \\
\textit{University of Hamburg}\\
Hamburg, Germany}
\and
\IEEEauthorblockN{Alina Mannanova \orcidlink{0009-0001-1792-1389}}
\IEEEauthorblockA{\textit{Signal Processing Group} \\
\textit{University of Hamburg}\\
Hamburg, Germany}
\and
\IEEEauthorblockN{Timo Gerkmann \orcidlink{0000-0002-8678-4699}}
\IEEEauthorblockA{\textit{Signal Processing Group} \\
\textit{University of Hamburg}\\
Hamburg, Germany}
}

\maketitle

\begin{abstract}
Due to their robustness and flexibility, neural-driven beamformers are a popular choice for speech separation in challenging environments with a varying amount of simultaneous speakers alongside noise and reverberation.
Time-frequency masks and relative directions of the speakers regarding a fixed spatial grid can be used to estimate the beamformer's parameters.
To some degree, speaker-independence is achieved by ensuring a greater amount of spatial partitions than speech sources. 
In this work, we analyze how to encode both mask and positioning into such a grid to enable joint estimation of both quantities.
We propose \acl{mwslc} and show that it achieves considerable performance in both tasks compared to baseline encodings optimized for either localization or mask estimation. %
In the same setup, we demonstrate superiority for joint estimation of both quantities. 
Conclusively, we propose a universal approach which can replace an upstream \acl{ssl} system solely by adapting the training framework, making it highly relevant in performance-critical scenarios.
\end{abstract}

\begin{IEEEkeywords}
Multi-channel, speech separation, sound source localization, mask estimation, speaker-independent
\end{IEEEkeywords}

\section{Introduction}
\label{sec:intro}
Given an audio recording containing multiple speakers, the field of \textit{speech separation} deals with the simultaneous disentanglement of all speech sources into individual audio streams.
While neural approaches have gained great proficiency in this task over the recent years, highly dynamic environments such as the so-called cocktail-party scenario \cite{cherry53cocktail_party} pose a great challenge until today.
On top of noisy and reverberated speech signals, the varying amount of simultaneous speakers makes this problem especially demanding.
In case recordings from a microphone array are available, the embedded spatial information can be leveraged to simplify the separation process.
Especially the non-linear integration of \textit{spatial} correlations and \textit{temporal-spectral} patterns between the multi-channel signals from the array leads to powerful and computationally efficient \ac{nn} architectures \cite{tesch24separation, quan24spatialnet, briegleb23icospa, yang23mcnet}. 

While popular concepts like \ac{pit} \cite{yu17pit} or its multi-channel extension \ac{lbt} \cite{taherian22sep_lbt} provide an efficient training framework, they are conceptually limited to a predetermined maximum amount of simultaneous
speech sources.
This dependency arises from the fixed output size, which is deeply rooted in the architecture of the \ac{nn}.
While iterative separation approaches can circumvent this issue by extracting each speaker at a time in a spatially guided fashion \cite{tesch24separation, briegleb23icospa, meng22lspex, nakagome20guided_sep, gu19neural_spatial_filter, pertila15loc_guided_sep}, the amount of subsequent \ac{nn} executions in scenarios with many participants limits their applicability \cite{bohlender24sep_journal}.
Neural \ac{ssl} systems \cite{desai22review_ssl} tackle the closely related problem of \textit{speaker-independent} localization by assuming that all speakers are uniquely identifiable through their relative positioning towards the microphone array.
By partitioning the recording environment into a fixed spatial grid, such approaches conduct either a binary or probabilistic speaker activity estimation for each individual region \cite{xiong15gcc_phat, weipeng18gcc_phat}.
Based on the success in \ac{ssl}, this method has been adapted for speech separation.
Instead of an activity indication, \ac{mwsbc} \textit{encodes} a time-frequency mask into each partition containing a speaker \cite{kindt22sep, hasfati22mwsbc, bohlender21lbt_sep, chazan19mwsbc}.
However, in the case of a very fine grid and sparse speech masks, there arises a significant label imbalance leading to an ill-conditioned optimization problem during training with regression-based loss functions \cite{kindt22sep, bohlender21lbt_sep}.
To avoid this issue, an alternative is to only utilize the output channels corresponding to the positions of the speakers. 
While this significantly improves the conditioning, it alleviates the need for the \ac{nn} to conduct precise localization, thus a separate \ac{ssl} system becomes necessary \cite{bohlender21lbt_sep}.

This work proposes a mask encoding which reduces the impact of label imbalance while at the same time retaining localization capabilities.
By spectrally enriching the \ac{slc} introduced in \cite{weipeng18gcc_phat}, we present \ac{mwslc} as alternative to \ac{mwsbc} and prove that it leads to a superior conditioning for a high spatial resolution.
Finally, we show that \ac{mwslc} dominates in \textit{joint} localization and mask estimation while retaining considerable performance in both tasks individually.

\section{Beamformer-guided Speech Separation}
\label{sec:problem}
Let the speech mixture $\mathbf{Y}$ contain $\channel$ microphone recordings of $I$ stationary speech sources $S^{\scriptscriptstyle(i)}$ in a reverberant but noiseless environment.
The signal propagation between $i$-th source and all microphones shall be represented by the \ac{rir} $\mathbf{H}^{\scriptscriptstyle(i)}$. 
In the \ac{stft} domain at time frame $t$ and frequency bin $k$, the acoustic scenario is thus resembled by
\begin{equation}\label{eq:signal_model}
    \mathbf{Y}_{tk} = \displaystyle \sum_{i} \mathbf{H}^{\scriptscriptstyle(i)}_{tk}S^{\scriptscriptstyle(i)}_{tk} \, .
\end{equation}
Due to its robustness and straightforward parameterization, we choose the \ac{mvdr} beamformer \cite{vary06mvdr} for speech separation.
Assuming knowledge about the microphone array geometry, it can be specified through the steering vector $\mathbf{d}^{\scriptscriptstyle(i)}_{tk}$ and the covariance matrix $\mathbf{R}^{\scriptscriptstyle(i)}_k$. Dropping indices for visual clarity, the separated speech signals $\widehat{S}^{\scriptscriptstyle(i)}_{tk}$ are given by 
\begin{equation}\label{eq:mvdr}
    \widehat{S} = \frac{\mathbf{d}^{\mathrm{H}}\mathbf{R}^{-1}\mathbf{Y}}{\mathbf{d}^{\mathrm{H}}\mathbf{R}^{-1}\mathbf{d}} \, .
\end{equation}
 The steering vector $\mathbf{d}^{\scriptscriptstyle(i)}_{tk}$ is computed from the associated \ac{doa} $\theta^{\scriptscriptstyle(i)}$ assuming anechoic propagation between the microphones \cite{vary06mvdr}. The second unknown, the spatial covariance matrix $\mathbf{R}^{\scriptscriptstyle(i)}_k$ of the interference w.r.t. each speaker, is estimated by
\begin{equation}\label{eq:covariance}
    \mathbf{R}^{\scriptscriptstyle(i)}_k = \frac{1}{T} \displaystyle \sum_{t} (1 - M^{\scriptscriptstyle(i)}_{tk}) \mathbf{Y}_{tk} \mathbf{Y}^{\mathrm{H}}_{tk}
\end{equation}
in an interval of T frames with time-frequency masks $M^{\scriptscriptstyle(i)}_{tk}$.
For this purpose we utilize \acp{irm} \cite{yuan18irm, wang16irm}, which we threshold by $\mathcal{E}_M$ to account for time-frequency bins without speech
\begin{equation} \label{eq:irm}
    M^{\scriptscriptstyle(i)}_{tk} = 
    \begin{cases} 
        \dfrac{{\lvert S^{\scriptscriptstyle(i)}_{tk}\rvert}^2}{{\lvert S^{\scriptscriptstyle(1)}_{tk} \rvert}^2 + \cdots + {\lvert S^{\scriptscriptstyle(I)} _{tk}\rvert}^2} & \text{if \,} \lvert S^{\scriptscriptstyle(i)}_{tk}\rvert > 10^{\frac{\mathcal{E}_M}{20\,\mathrm{dB}}}\, ,\\ 
        0 & \text{else.}
    \end{cases}  
\end{equation} 
Consequently both \acp{doa} $\theta^{\scriptscriptstyle(i)}$ and \acp{irm} $M^{\scriptscriptstyle(i)}_{tk}$ are required for beamforming.
Joint estimation of these quantities is the focus of this work. 

\section{Mask-Weighted Spatial Coding}
\label{sec:lbo}

\subsection{Mask-Weighted Spatial Binary Coding}
\label{ssec:lboc}
In this section we revisit \acf{mwsbc} and analyze why regression-based loss functions lead to an ill-conditioned training \cite{kindt22sep, bohlender21lbt_sep, boeddeker24ts_sep}.
\ac{mwsbc} divides the space around the microphone array into $\Theta$ segments and assigns time-frequency masks to the directions $\theta^{\scriptscriptstyle(i)}$ where the speakers are present. The remaining segments are filled with zeros, so that the resulting encoding $L_{tk\theta}$ can be expressed as
\begin{equation}\label{eq:mwsbc}
    L_{tk\theta} = \displaystyle \sum_{i} M^{\scriptscriptstyle(i)}_{tk}\delta_{\theta \theta^{\scriptscriptstyle(i)}} \, ,
\end{equation}
with $\delta_{\theta \theta^{\scriptscriptstyle(i)}}$ denoting the Kronecker delta \cite{arens18mathematik}, a binary indicator function which is non-zero only if $\theta$ equals $\theta^{\scriptscriptstyle(i)}$.
This can be seen as spectral extension of the binary \ac{doa} encoding \cite{tesch24separation, bohlender21ssl_temporal_context, papageorgiou21ssl, chakrabarty19ssl} which we will denote by \ac{sbc}. 
In case a high spatial resolution is desired, the number of segments $\Theta$ is much greater
than the number of speakers $I$, thus, the majority of regions are empty. Additionally, due to the sparsity of speech signals in the time-frequency domain, the bins of the remaining partitions containing speakers are also predominantly inactive, which is why MW-SBC suffers from
a significant label imbalance.
Throughout the training of a \ac{nn}, the resulting vast amount of zeros has a significant impact on the gradient during backpropagation.
Especially regression-based loss functions, which are known to be robust against outliers, are prone to converge to poor, quasi-stationary solutions for the estimate $\widehat{L}_{tk\theta}$ of ${L}_{tk\theta}$, such as $\widehat{L}_{tk\theta} = 0$ \cite{boeddeker24ts_sep}.
Due to being a famous representative of regression-based loss functions, we will examine the \ac{mse} more closely.
Omitting the average about time and frequency bins, the \ac{mse} between $\widehat{L}_{tk\theta}$ and ${L}_{tk\theta}$ is defined by
\begin{align}
\mathcal{L}^{\mathrm{MSE}}_{tk} &= \frac{1}{\Theta} \displaystyle \sum_{\theta} {\left( L_{tk\theta} - \widehat{L}_{tk\theta} \right)}^2 \, . \\
\intertext{
Specifically, we are interested in the gradient regarding $\widehat{L}_{tk\theta}$,}
\dfrac{\partial \mathcal{L}^{\mathrm{MSE}}_{tk}}{\partial \widehat{L}_{tk\theta}} &= \frac{2}{\Theta} \left( \widehat{L}_{tk\theta} - L_{tk\theta}  \right) \, , \\
\intertext{as it weights all differentiations w.r.t. the \ac{nn} parameters according to the chain rule. 
Its $\mathrm{L}_1$ norm regarding the spatial dimension yields}
\left\lVert \dfrac{\partial \mathcal{L}^{\mathrm{MSE}}_{tk}}{\partial \widehat{\mathbf{L}}_{tk}} \right\rVert_1 &= \frac{2}{\Theta}  \displaystyle \sum_{\theta} \left| \widehat{L}_{tk\theta} - L_{tk\theta}  \right| \label{eq:mse_grad_norm} \, , 
\end{align}
\begin{align}
\intertext{with $\widehat{\mathbf{L}}_{tk}$ representing the spatial vectorization of $\widehat{L}_{tk\theta}$.
Examining the norm at $\widehat{L}_{tk\theta}=0$ and inserting \eqref{eq:mwsbc} gives}
    \left\lVert \dfrac{\partial \mathcal{L}^{\mathrm{MSE}}_{tk}}{\partial \widehat{\mathbf{L}}_{tk}} \right\rVert_{1,\widehat{L}_{tk\theta}=0} &= \frac{2}{\Theta} \displaystyle \sum_{i} M^{\scriptscriptstyle(i)}_{tk} \, . \\
\intertext{Since the \acp{irm} \eqref{eq:irm} are bounded by 1, the expression approaches zero in the limiting case of infinitely many partitions $\Theta$}
    \lim_{\Theta \to \infty} \left\lVert \dfrac{\partial \mathcal{L}^{\mathrm{MSE}}_{tk}}{\partial \widehat{\mathbf{L}}_{tk}} \right\rVert_{1,\widehat{L}_{tk\theta}=0} &= 0 \label{eq:grad_mwsbc_0} \, .
\end{align}
Assuming the remaining differential chain is bounded, the sub-multiplicative property of norms \cite{bjoerk96numeric} leads to a vanishing of the whole gradient w.r.t. the \ac{nn} parameters.
In other words, with a very large amount of segments $\Theta$, $\widehat{L}_{tk\theta}=0$ is correct in almost all cases. The remaining non-zero partitions containing \acp{irm} have only negligible influence, which leads to a plateau in training.
To avoid this problem, \cite{kindt22sep, bohlender21lbt_sep} solely utilize the masks of active speakers during loss computation.
While this removes the majority of zeros, it also decouples precise localization from the training objectives.

\subsection{Proposed Mask-Weighted Spatial Likelihood Coding}
\label{ssec:lbol}
In this subsection we present our proposed method to overcome the vanishing gradient problem of \ac{mwsbc} \cite{bohlender21lbt_sep} and at the same time extend it towards \textit{joint} localization and mask estimation.
Our approach is inspired by the \acf{slc} introduced in \cite{weipeng18gcc_phat} within the context of \ac{ssl}, which has gained a lot of popularity in recent \ac{ssl} systems \cite{meng22lspex, yin24mimo_doa, fu22issl, nguyen20ssl_slc}. 
The idea of \ac{slc} is to replace \ac{sbc} by Gaussian curves centered at the \acp{doa}.
We propose to enrich this spatial likelihood 
with the spectral information from the \acp{irm} $M^{\scriptscriptstyle(i)}_{tk}$ by weighting the Gaussians accordingly.
The resulting spatio-spectral coding, which we denote as \acf{mwslc}, is defined as
\begin{equation}\label{eq:mwslc}
    L_{tk\theta} = \underset{i \in I}{\mathrm{max}} \ M^{\scriptscriptstyle(i)}_{tk} \, e^{-d{\left( \theta, \theta^{\scriptscriptstyle(i)} \right) }^2\big/\sigma^2}
\end{equation}
with the standard deviation $\sigma$ of the Gaussians.
The angular distance $d(\cdot,\!\cdot)$ represents the wrapped \ac{mae} \cite{feng23quantization}, thus taking the circularity of the \acp{doa} into account.
Similar to \cite{weipeng18gcc_phat}, we accumulate the individual curves with a maximum operation.
However, in our case, due to weighting with the \acp{irm}, the peaks of the underlying curves are no longer equally prominent.
Especially for small angular distances and frequency bins dominated by a single speaker, adjacent masks can influence each other in our coding.
While a $\sigma$ chosen in correspondence to a minimum spatial gap between the speakers can alleviate this effect, it is a conceptual limitation \ac{mwsbc} does not suffer.
To enable further analysis, we approximate the non-linear maximum operator in \eqref{eq:mwslc} with a sum 
\begin{equation}
    L_{tk\theta} \approx \displaystyle \sum_i M^{\scriptscriptstyle(i)}_{tk} \, e^{-d{\left( \theta, \theta^{\scriptscriptstyle(i)} \right) }^2\big/\sigma^2} \, ,
\end{equation}
which is valid under the assumption of Gaussians with low standard deviations and large mean difference.
Based on this approximation, we conduct the same investigation for the \ac{mse} gradient norm at $\widehat{L}_{tk\theta}=0$ as for \ac{mwsbc}.
Combining \eqref{eq:mse_grad_norm} with \eqref{eq:mwslc} at $\widehat{L}_{tk\theta}=0$ gives
\begin{equation}
\left\lVert \dfrac{\partial \mathcal{L}^{\mathrm{MSE}}_{tk}}{\partial \widehat{\mathbf{L}}_{tk}} \right\rVert_{1,\widehat{L}_{tk\theta}=0} \approx \frac{2}{\Theta} \displaystyle \sum_{\theta} \sum_{i} M_{tk}^{\scriptscriptstyle(i)} e^{-d{\left( \theta, \theta^{\scriptscriptstyle(i)} \right) }^2\big/\sigma^2} \, . 
\end{equation}
We rewrite this expression by introducing the constant $\Omega$
\begin{equation}
\left\lVert \dfrac{\partial \mathcal{L}^{\mathrm{MSE}}_{tk}}{\partial \widehat{\mathbf{L}}_{tk}} \right\rVert_{1,\widehat{L}_{tk\theta}=0} = \frac{2}{\Omega} \displaystyle \sum_{i}  M_{tk}^{\scriptscriptstyle(i)} \sum_{\theta}  e^{-d{\left( \theta, \theta^{\scriptscriptstyle(i)} \right) }^2\big/\sigma^2} \,  \frac{\Omega}{\Theta} \, ,
\end{equation}
with $\Omega$ representing the angular space around the array in which the speakers can be located, e.g. in the case of unconstrained speaker positioning $\Omega=360^\circ$.
Considering the limiting case of a very high spatial resolution, thus the amount of segments $\Theta$ approaches infinity, the partition size $\frac{\Omega}{\Theta}$ transitions to the differential $d\theta$.
Together with the sum over the infinite amount of segments, the limit converges to the integral expression
\begin{align}
\lim_{\Theta \to \infty} &\left\lVert \dfrac{\partial \mathcal{L}^{\mathrm{MSE}}_{tk}}{\partial \widehat{\mathbf{L}}_{tk}} \right\rVert_{1,\widehat{L}_{tk\theta}=0} \notag \\
&\hspace*{40pt}\approx \frac{2}{\Omega} \displaystyle \sum_{i} M_{tk}^{\scriptscriptstyle(i)} \int_\Omega e^{-{d\left( \theta,\theta^{\scriptscriptstyle(i)} \right) }^2\big/\sigma^2} d\theta \, .\\
\intertext{
Since the Gaussian function is strictly positive, the integral also evaluates to a positive non-zero value.
Therefore, in contrast to \ac{mwsbc} in \eqref{eq:grad_mwsbc_0}, the gradient's norm of our proposed \ac{mwslc} depends on the \acp{irm} $M_{tk}^{\scriptscriptstyle(i)}$ and is non-zero in speech presence.
The main difference between both approaches lies in the continuous nature of the Gaussian encoding, 
which increases the spatial prominence with a rising amount of partitions $\Theta$, while it stays constant for the Kronecker delta function used in \eqref{eq:mwsbc}.
\endgraf 
To find a closed-form solution, we assume an unconstrained speaker positioning.
By choosing the lower (equals upper) limit of integration centered between two \acp{doa} $\theta^{\scriptscriptstyle(i)}$, the circularity of the angular distance becomes negligible due to the assumption of Gaussians with low standard deviations and large mean difference.
In the same manner, the integration can be extended to the whole real number line, yielding
}
&\hspace*{40pt}\approx \frac{2}{\Omega} \displaystyle \sum_{i} M_{tk}^{\scriptscriptstyle(i)} \int\limits_{-\infty}^\infty e^{-{\left| \theta-\theta^{\scriptscriptstyle(i)} \right| }^2\big/\sigma^2} d\theta \notag \\
&\hspace*{40pt}= \sqrt{\pi}\,\frac{2\sigma}{\Omega} \displaystyle \sum_{i} M_{tk}^{\scriptscriptstyle(i)} \, .
\end{align}
The norm's proportionality to $\sigma$ underlines the strong connection between problem conditioning and standard deviation $\sigma$ of the Gaussians. 
While an increased $\sigma$ improves training, it also puts less weight on accurate localization and possibly results in a reduced mask quality.

\subsection{Joint DoA and Mask Extraction}
\label{ssec:joint_est}
\begin{figure}[t!]
    \centering
   \input{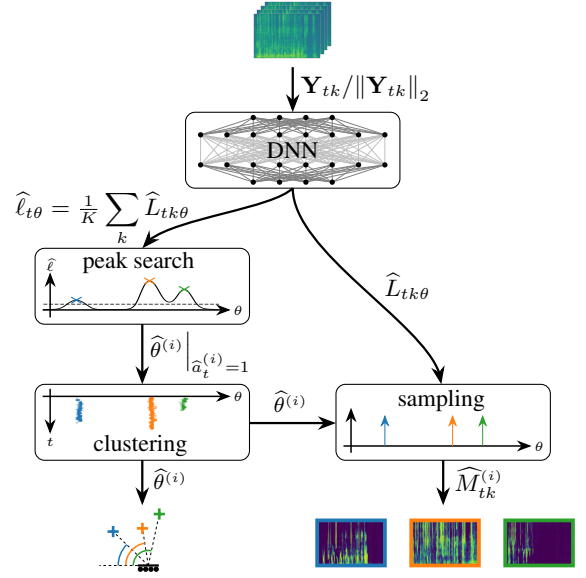}
    \caption{Joint localization and mask estimation framework. The estimated spatio-spectral coding $\widehat{L}_{tk\theta}$ is averaged and post-processed to obtain the time-independent \ac{doa}s $\widehat{\theta}^{\scriptscriptstyle(i)}$. Finally, the \acp{irm} $\widehat{M}_{tk}^{\scriptscriptstyle(i)}$ are recovered by sampling $\widehat{L}_{tk\theta}$ at $\widehat{\theta}^{\scriptscriptstyle(i)}$.}
    \label{fig:system_overview}
\end{figure}

\Cref{fig:system_overview} summarizes the process of predicting $\widehat{L}_{tk\theta}$ with a \ac{dnn} and its application to both mask and \ac{doa} estimation.
The latter is indicated on the left, where a \textit{peak-search} w.r.t. $\widehat{\ell}_{t\theta}$ is employed, the average of $\widehat{L}_{tk\theta}$ regarding all frequency bins $K$,
\begin{equation}\label{eq:sslc_peaks}
     \widehat{\theta}^{\scriptscriptstyle(i)} \Big|_{\widehat{a}^{\scriptscriptstyle(i)}_t=1} = \underset{\mathcal{E}_\theta, \Delta\theta}{\mathrm{peakpos}} \ \widehat{\ell}_{t\theta}\, ,
\end{equation}
returning all angles at which $\widehat{\ell}_{t\theta}$ is above a threshold $\mathcal{E}_\theta$ and prominent in a neighborhood $\pm \Delta \theta$.
Due to the frame-wise estimation, only the \ac{doa}s of the locally active speakers are found.
This is represented by the estimated binary speech activity indicator $\widehat{a}^{\scriptscriptstyle(i)}_t$.
Therefore, to identify all \ac{doa}s $\widehat{\theta}^{\scriptscriptstyle(i)}$ in the recording, a post-processing step e.g. in form of \textit{clustering} has to be applied.
Finally, as indicated on the right-hand side of \cref{fig:system_overview}, the corresponding masks are extracted by slicing the estimated coding according to the \ac{doa} estimates, which we denote as \textit{sampling}
\begin{equation}\label{eq:sslc_masks}
    \widehat{M}_{tk}^{\scriptscriptstyle(i)} = \widehat{L}_{tk{\widehat{\theta}^{\scriptscriptstyle(i)}}} \, .
\end{equation}

\begin{table*}[t]
\resizebox{\textwidth}{!}{
\renewcommand{\arraystretch}{0.9}
\footnotesize
\begin{tabular}{clllccccccc}
 \toprule[1.5pt]
 & \multicolumn{3}{c}{\textbf{Model}} & \multicolumn{2}{c}{\textbf{Estimation}} & \multicolumn{3}{c}{\textbf{Localization}} & \multicolumn{2}{c}{\textbf{Separation}} \\[1pt] \cmidrule(lr){2-4} \cmidrule(lr){5-6} \cmidrule(lr){7-9} \cmidrule(lr){10-11} 
\textbf{ID} & Encoding & Architecture & Loss & DoA & IRM & MAE\,($^{\circ}$)\,$\downarrow$ & Precision\,(\%)\,$\uparrow$ & Recall\,(\%)\,$\uparrow$ & $\Delta$\acs{sisdr}\,(dB)\,$\uparrow$ & \acs{estoi}\,(\%)\,$\uparrow$  \\ \midrule
(1) & - & Oracle & - &  \textcolor{red}{\ding{55}} & \textcolor{red}{\ding{55}} & -\,/\,- & -\,/\,- & -\,/\,- & 6.60\,/\,7.33 & 69.2\,/\,57.9 \\ [-1pt] \cmidrule(lr){1-11}
(2) & \acs{sbc} & CNN/LSTM\cite{bohlender21ssl_temporal_context} & \acs{bce} & \textcolor{green}{\ding{51}} & - & 3.33\,/\,6.82 & 81.4\,/\,84.7 & 94.7\,/\,75.2 & -\,/\,- & -\,/\,- \\
(3) & \acs{slc} & CNN/LSTM\cite{bohlender21ssl_temporal_context} & \acs{mse} & \textcolor{green}{\ding{51}} & - &\textbf{0.63}\,/\,\textbf{2.36} & \textbf{93.3}\,/\,\textbf{94.0} & \textbf{95.2}\,/\,\textbf{83.3} & -\,/\,- & -\,/\,- \\[-1pt] \cmidrule(lr){1-11} %
(4) &  \acs{mwsbc} & FB-MEst\cite{bohlender21lbt_sep} & \acs{mse}$^\dagger$ & \textcolor{red}{\ding{55}} & \textcolor{green}{\ding{51}} & -\,/\,- & -\,/\,- & -\,/\,- & 5.26\,/\,6.38 & 63.6\,/\,54.5\\
(5) & \acs{mwsbc} & MC-CRUSE\cite{kindt22sep} &  \acs{mse}$^\dagger$  & \textcolor{red}{\ding{55}} & \textcolor{green}{\ding{51}} & -\,/\,- & -\,/\,- & -\,/\,- & \textbf{5.85}\,/\,\textbf{6.53} & \textbf{65.5}\,/\,\textbf{55.0}\\ [-1pt] \cmidrule(lr){1-11}
(6) & \acs{mwsbc} & FB-MEst\cite{bohlender21lbt_sep} & \acs{mse}$^\dagger$ & \textcolor{green}{\ding{51}} & \textcolor{green}{\ding{51}} & 20.74\,/\,28.53 & 52.0\,/\,39.3 & 60.3\,/\,55.5 & \hspace{0.7pt} 3.08\,/\,-0.25 & 60.8\,/\,48.5 \\
(7) &  \acs{mwslc}\,(ours) & FB-MEst & \acs{mse} & \textcolor{green}{\ding{51}} & \textcolor{green}{\ding{51}} & \textbf{1.26}\,/\,\textbf{3.05} & \textbf{89.4}\,/\,\textbf{90.1} & \textbf{94.5}\,/\,\textbf{85.4} & 4.99\,/\,\textbf{5.92} & \textbf{62.8}\,/\,\textbf{53.0} \\
(8) & \acs{mwsbc} & MC-CRUSE\cite{kindt22sep} & \acs{mse}$^\dagger$  & \textcolor{green}{\ding{51}} & \textcolor{green}{\ding{51}} & 20.65\,/\,28.61 & 45.0\,/\,36.8 & 48.6\,/\,44.5 & \hspace{0.7pt} 0.26\,/\,-1.54 & 59.9\,/\,48.8\\
(9) &  \acs{mwslc}\,(ours) & MC-CRUSE & \acs{mse} & \textcolor{green}{\ding{51}} & \textcolor{green}{\ding{51}} & 2.32\,/\,7.24 & 88.8\,/\,88.4 & 92.1\,/\,77.6 & \textbf{5.03}\,/\,5.64 & 62.6\,/\,52.3\\ 
 \bottomrule[1.5pt]
\end{tabular}
}
 \caption{
 Localization and separation performance on MC-Libri2Mix\,/\,MC-Libri3Mix datasets. The input \acs{sisdr} and \acs{estoi} scores are -5.46\,/\,-8.14\,dB and 46.2\,/\,34.4\,\% respectively.
 $^\dagger$ indicates that during training only the masks of active speakers are considered.
}
\label{tab:results}
\end{table*}

\section{Experiments}
\label{sec:experiments}
\subsection{Experimental Setup}
\label{ssec:setup}
\textit{\textbf{Dataset}} For training and evaluation we utilize the publicly available multi-channel dataset MC-LibriMix \cite{meng22lspex}. 
Specifically, we employ the versions with a sampling rate of 16-kHz
generated for two (MC-Libri2Mix) and three (MC-Libri3Mix) speakers.
The spatialization was conducted with simulated RIRs for a linear 4-channel
microphone array. The simulations are constrained to reverberation times between 200\,ms and 600\,ms and a minimum angular distance between adjacent speakers of 15\,$^\circ$ (more details in [8]). Regarding the STFT, we use a square-root Hann window \cite{shimauchi14hann_window} of length 32\,ms and 16\,ms hop-size.

\indent \textit{\textbf{DoA and Mask Estimation}}
The standard deviation $\sigma$ of the Gaussian curves in both \ac{slc} and \ac{mwslc} is set to 6\,$^\circ$ as in \cite{meng22lspex}. The same value is used for the neighborhood $\pm\Delta\theta$ in \eqref{eq:sslc_peaks} as proposed in \cite{weipeng18gcc_phat}.
The threshold $\mathcal{E}_\theta$ is determined via an exhaustive search on the validation dataset of MC-Libri2Mix w.r.t. optimizing the $F_1$ score.
Finally, we utilize \ac{hac} to first obtain the \acp{doa} and then the \acp{irm} for all identified speakers.
In each step we use the average distance between separate clusters as linkage and set $2\sigma$ as merging-condition.
The threshold $\mathcal{E}_M$ for the \ac{irm} computation in \eqref{eq:irm} is set to -35\,dB.

\subsection{Network Architectures and Training Details}
\label{ssec:training}
\textit{\textbf{\acs{nn} Architecture}}
To evaluate our proposed encoding, we employ two \ac{nn} architectures introduced for \ac{mwsbc}, namely \acs{fbmest} \cite{bohlender21lbt_sep} and \acs{mccruse} \cite{kindt22sep}.
As a reference, we also list the \ac{ssl} model CNN/LSTM \cite{bohlender21ssl_temporal_context} due to its close relation to \acs{fbmest}.
To further improve comparability, we utilize the mixture $\mathbf{Y}$ normalized regarding the channel dimension as input features \cite{kindt22sep, bohlender21lbt_sep}, see \cref{fig:system_overview}, and equip all architectures with Sigmoid output activations.

\textit{\textbf{Training}} 
All \ac{nn}s are trained on MC-Libri2Mix with a batch-size of 5 for 100 epochs or until convergence, which we define as no improvement on the validation dataset for 10 consecutive epochs.
MC-Libri3Mix is only used during evaluation.
The initial learning rate is set to 0.001 and reduced by a factor of 0.63 every 10 epochs, thus, leading to a decimation approximately every 50 epochs.

\section{Results}
\label{ssec:performance}
\Cref{tab:results} displays the results regarding \ac{ssl} and joint localization and mask estimation in terms of separation performance with an \ac{mvdr} beamformer. 
The loss computation is indicated by columns \textit{Encoding} and \textit{Loss} with \ac{sbc} and \ac{slc} following \cite{bohlender21ssl_temporal_context} and \cite{weipeng18gcc_phat} respectively.

\begin{figure}[t]
  \centering
   \input{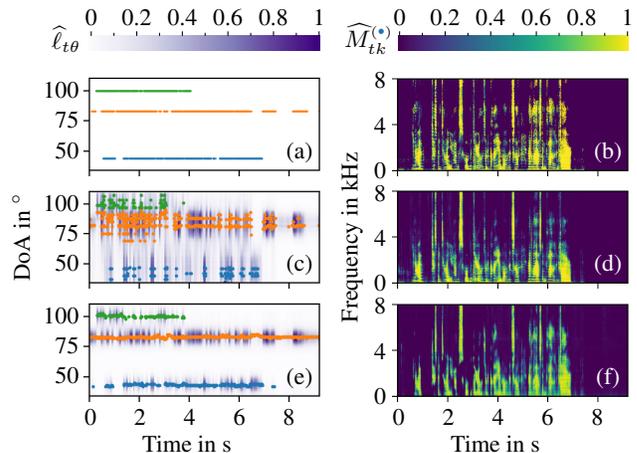}
   \caption{Ground truth (a,\,b), estimated \ac{mwsbc} (c,\,d) and our proposed \ac{mwslc} (e,\,f) with \acs{mccruse} from top to bottom. Averaged coding $\widehat{\ell}_{t\theta}$ with clustered \ac{doa}s and corresponding masks $\widehat{M}^{\scriptscriptstyle(i)}_{tk}$ for speaker \mbox{(\!
   \begin{tikzpicture}[baseline]
        \fill[tab_blue] (0,0.09) circle (1pt); %
    \end{tikzpicture}
    \!)} on the left and right respectively. The mask for \ac{mwsbc} (d) is extracted using oracle \acp{doa}. 
    }
   \label{fig:likelihood_spectogram}
\end{figure}

\textit{\textbf{Localization}} 
As it is common procedure in \ac{ssl}, we assess the performance both in terms of a known (\acs{mae}) and unknown (precision, recall) amount of speakers according to the recipe from \cite{weipeng18gcc_phat}.
From two to three speakers almost all methods demonstrate increased precision and decreased recall scores.
This can be traced back to optimizing the peak-search for the two speaker case, as the threshold $\mathcal{E}_\theta$ is set too high to ideally accommodate more speakers.
Regarding the baseline CNN/LSTM architecture, employing \ac{slc} as output coding (3) outperforms \ac{sbc} (2) by a large margin.
Possibly due to their similar \ac{nn} architectures, \acs{fbmest} trained with \ac{mwslc} achieves almost the same localization results in (7).
On the other hand, \acs{mccruse}, which is originally not affiliated with \ac{ssl} (5), displays slightly inferior performance (9).
As expected, both \acp{nn} trained with \ac{mwsbc} (6,\,8) perform significantly worse than with our proposed encoding in (7,\,9), which can be also seen from a comparison between the \ac{doa} estimates in \cref{fig:likelihood_spectogram} (c) and (e).

\textit{\textbf{Separation}} To evaluate speaker-independent separation performance, we follow the recipe proposed in \cite{higuchi17deep_clustering}. 
The general gap between separation with an oracle \ac{irm} (1) and an estimated mask by employing \ac{mwsbc} (4,\,5) can be attributed to a low input \ac{snr} of -5.50\,dB for two and \mbox{-8.18\,dB} for three speakers.
Furthermore, the considerable difference between \acs{fbmest} (4) and \acs{mccruse} (5) can be linked to the architecture of \acs{fbmest},
which leads to masks lacking the fine harmonic structure of speech, see \cite{bohlender21lbt_sep}.
Since the explicit localization criterion is dropped during training, \acs{mwsbc} supplied with oracle \ac{doa}s (4,\,5) represents a natural upper bound for our proposed \acs{mwslc} (7,\,9).
Regarding \acs{fbmest} (7), both objective and perceptive measures come very close to this bound (4), however, the performance gap for \acs{mccruse} (9) is more pronounced.
As seen in \cref{fig:likelihood_spectogram}, \acs{mwslc} (f) does not match the temporal-spectral detail \acs{mwsbc} (d) achieves.
We propose that this shortcoming could be improved by skewing the \acs{mwslc} training loss with an additional term purely for mask reconstruction in future work. 
For completeness, we have also listed the separation results using estimated \ac{doa}s from \acs{mwsbc} (6,\,8), which are clearly outperformed by our proposed method (7,\,9).

\section{Conclusion}
\label{sec:conclusion}
In this work we proposed \acf{mwslc} for speaker-independent joint localization and mask estimation.
Backed by theoretical investigations, we evaluated our approach towards encodings optimized for either localization or mask estimation.
We showed that our method achieves considerable performance in both tasks w.r.t. the baselines, although performing them conjointly with equal computational overhead.
In the same setup we demonstrated unmatched dominance for joint estimation. 
Conclusively, we proposed a universal approach which replaces an upstream \ac{ssl} system by simply adapting the training scheme, making it highly relevant in performance-critical scenarios.

\bibliographystyle{IEEEbib_initials}
\bibliography{strings,refs}
\end{document}